\documentclass{article}
\usepackage{graphicx,color,amsmath}
\usepackage[top=20truemm,bottom=30truemm,left=25truemm,right=25truemm]{geometry}

\newcommand{\Slash}[1]{\ooalign{\hfil/\hfil\crcr$#1$}}
\newcommand{\pa}{\partial}
\newcommand{\nn}{\nonumber}

\newcommand{\bref}[1]{(\ref{#1})}

\newcommand{\al}{\alpha}

\newcommand{\Gammaeff}{\Gamma_{I}}

\begin{document}
\title{\bf QED in the Exact Renormalization Group}
\author{Yuji Igarashi and Katsumi Itoh\\
\hfill\\
Faculty of Education, Niigata University,\\ Niigata 950-2181, Japan\\
\hfill\\
igarashi@ed.niigata-u.ac.jp\\
itoh@ed.niigata-u.ac.jp
}

\maketitle

\begin{abstract}%
The functional flow equation and the Quantum Master
equation are consistently solved in perturbation for the chiral
symmetric QED with and without four-fermi interactions.  Due to the
presence of momentum cutoff, unconventional features related to gauge
symmetry are observed even in our perturbative results. 

In the absence of the four-fermi couplings, one-loop calculation gives
us the Ward identity, $Z_{1}=Z_{2}$, and the standard results of
anomalous dimensions and the beta function for the gauge coupling.
It is a consequence of regularization scheme independence in
one-loop computation.  We also find a photon mass term.

When included, four-fermi couplings contribute to the beta
function and the Ward identity is also modified, $Z_{1}
\neq Z_{2}$, due to a term proportional to the photon mass multiplied
by the four-fermi couplings.
\end{abstract}


\section{Introduction}

Recently much attention has been attracted to the exact renormalization
group or functional renormalization group (ERG/FRG) approach to gauge
theories. The regularization scheme with a momentum cutoff $\Lambda$ is
not compatible with gauge invariance: the BRST transformation in its
standard form is not a symmetry of the Wilsonian action.  However it has
been shown that the BRST symmetry survives in a modified form
\cite{Igarashi:1999rm,Sonoda:2007dj,Igarashi:2008bb,Igarashi:2009tj}:
the variation of a Wilsonian action $S$ under appropriately modified
BRST transformation defined at $\Lambda$ is cancelled by the Jacobian
factor of the functional measure.  This cancellation mechanism, the
modified Ward-Takahashi (mWT) identity, is lifted to the Quantum Master
Equation (QME) 
\cite{Igarashi:2000vf,Igarashi:2001mf,Igarashi:2007fw,Higashi:2007ax}
in the Batalin-Vilkovisky (BV) antifield formalism
\cite{BATALIN198127}.
The QME and the flow equation are two basic equations to define a gauge
 theory in ERG/FRG. It has been a challenging problem to solve them
 consistently in appropriate truncation schemes.

In a previous work \cite{10.1093/ptep/ptz099}, the compatibility of two
equations is discussed for Yang-Mills (YM) theory in a perturbative
framework (see also \cite{10.21468/SciPostPhys.5.4.040}). The main
results obtained there are: firstly, two equations are combined to
develop BRST cohomology analysis 
\cite{Fisch:1990uj,Henneaux:1991vp,Barnich:1995wt,Barnich:1995wm}
that uniquely determines the classical
action of first and second orders in gauge coupling; secondly, it was
shown that one-loop perturbative solution to the flow equation satisfies the QME
or its Legendre transform, modified Slavnov-Taylor (mST) identity
\cite{Ellwanger:1994iz}, up to third order in the coupling; thirdly the
standard results are obtained for the beta function and anomalous
dimensions as a consequence of regularization scheme independent
computation. It leads to the standard Slavnov-Taylor identities among
renormalization constants.

In this paper, we consider the compatibility between the QME and the
flow equation for a chiral invariant QED with four-fermi interactions
\cite{1751-8121-49-40-405401}.  This type of the model have attracted
interests in connection with possible existence of a non-trivial
UV-fixed point and associated chiral symmetry breaking
\cite{Maskawa:1974vs,10.1143/PTP.54.860,Fukuda:1976zb,Miransky:1984ef,PhysRevD.39.2430,PhysRevLett.56.1230,LEUNG1986649,Aoki:1996fh}.
In the light of asymptotically safe scenario, there is a new interest of
finding a UV completion of QED \cite{Gies2020}.  In this paper we will
not discuss such non-perturbative structure. Instead, we take a
perturbative approach in parallel to Ref.\cite{10.1093/ptep/ptz099} and
find how the higher dimensional interactions affect the realization of
BRST symmetry.

Here we mainly use the Legendre transform of the QME and flow equation
to avoid redundancy arising from the one-particle reducible part of
the Wilsonian action. QME/mST is also best expressed in terms of $\Gamma$ since its free
part $\Gamma_{0}$ carries no regularization that simplifies BRST
cohomology analysis.  Though the Legendre transform of the measure
contribution $\Delta S$ in QME contains the inverse of the two-point
function $\Gamma^{(2)}$, that is readily expandable perturbatively and
does not cause any trouble.

We will show that, even in the presence of the four-fermi interactions, 
the perturbative solution to the flow equation satisfies the QME/mST 
up to order of $e^{3}$ or $e G_{S,V}$ in a general covariant gauge. 

After introducing the wavefunction
renormalization factors via a canonical transformation of classical
fields and their antifields, the beta function of the gauge coupling
and anomalous dimensions are computed by using the flow equations.

Without four-fermi couplings, the standard perturbative results are
obtained as in the case of YM theory and the Ward identity $Z_{1} =
Z_{2}$ holds.  It is again a consequence of regularization scheme
independence.  The presence of the photon mass term proportional to $e^2
\Lambda^2$ is observed.  Once the four-fermi interactions are taken into
accounts, the beta function acquires an extra term proportional to the
photon mass term and we find $Z_{1} \ne Z_{2}$ due to the mass term.  
Still, our perturbative solution satisfies the
QME/mST by including the photon mass and the modified Ward identity
among $Z_{1}$ and $Z_{2}$.  We emphasize the fact that in ERG/FRG the
BRST symmetry is realized in a modified form.
 
In the next section, we give a brief summary of the 1PI formulation in
ERG/FRG.  In section 3, we show that one-loop perturbative solution to
the flow equation satisfies the QME/mST up to the third order in
couplings. The beta function and anomalous dimensions are computed in
section 4.  Summary and conclusions are given in section 5.

 \section{Legendre transform of the QME and the flow equation}

The Wilsonian action consists of free and interaction parts, $S=S_{0} + S_{I}$.  In the free action 
\begin{eqnarray}
&& 
S_{0}[\phi,\phi^{*}] = \frac{1}{2} \phi^A K^{-1} \Delta^{-1}_{AB}\,\phi^B 
+\phi^*_A K^{-1} R_{\ B}^A\phi^B
\,,
\label{S_0}
\end{eqnarray} 
the kinetic terms $\Delta^{-1}_{AB}$ are regularized by a UV cutoff
function $K(p^2/\Lambda^2)$, satisfying the requirements that $K(0)=1$
and $K(u)\to0$ sufficiently fast as $u\to\infty$. Also included are free
BRST transformations $R_{\ B}^A\phi^B$ for fields $\phi^{A}$ multiplied
by their antifields $\phi^{*}_{A}$ and by overall factor $K^{-1}$.
 By construction, the free BRST transformation satisfies the relation
\begin{eqnarray}
\Delta^{-1}_{AC}R^C_{\ B} + \Delta^{-1}_{BC}R^C_{\ A} =0
\label{Delta R relation}
\end{eqnarray}
$S_{I}[\phi,\phi^{*}]$
consists of interaction terms and some antifield dependent terms with
coupling constants.  We use the condensed notation as in
Refs.\cite{Igarashi:2009tj}\cite{10.1093/ptep/ptz099}.

The regularized version of the antibracket and the measure operator can
be defined as those in \cite{10.1093/ptep/ptz099}:
\begin{eqnarray}
(X,Y)_{K} = \frac{\partial^{r} X}{\partial\phi^A}\,
K\frac{\partial^{l}Y}{\partial\phi^*_A}-\frac{\partial^{r} X}{\partial\phi^*_A}\,
K\frac{\partial^{l}Y}{\partial\phi^A}~,
\label{antibracket}
\end{eqnarray}
and 
\begin{eqnarray}
\Delta_{K} X = (-)^{A+1} \frac{\partial^{r}}{\partial\phi^A}\,K\frac{\partial^{r}}{\partial\phi^*_A}X\,.    
\label{AB}
\end{eqnarray}
Here $X$ and $Y$ are arbitrary bosonic or fermionic functionals, and
$(-)^A=(-)^{\epsilon_A}$ where $\epsilon_A$ is the Grassmann parity of
$\phi^A$. $\phi^*_A$ has the opposite Grassmann parity to $\phi^A$.
$\pa^{l(r)}$ denotes the left(right) derivative.
The BRST invariance of the Wilsonian action is expressed as the QME on the fields and their antifields
\begin{eqnarray}
\Sigma = \frac{1}{2}(S,S)_{K} -\Delta_{K} S =0\,.
\label{QME_S}
\end{eqnarray}

The Wilsonian action $S$ can be expressed as a tree-level expansion in
terms of its 1PI part $\Gamma_{I}$ \cite{Morris:1993qb,Ishikake:2005rk,Morris:2015tg}.  The latter is
related to $S_{I}[\phi,\phi^{*}]$ via Legendre transformation
\begin{eqnarray}
&&\Gamma_I[\Phi,\Phi^*] = S_I[\phi,\phi^*] - \frac{1}{2}\, (\phi -\Phi)^A\, 
\bar\Delta^{-1}_{ AB}\,(\phi-\Phi)^B\,,
\label{Legendre}
\\
&&\frac{\partial^r}{\partial\phi^B} S_I[\phi,\phi^*] =  (\phi-\Phi)^A 
\bar\Delta^{-1}_{AB} = 
\frac{\partial^r}{\partial\Phi^B} \Gamma_I[\Phi,\Phi^*]\,,
\label{Legendre-1}
\end{eqnarray}
where $\bar\Delta^{-1}_{AB}$ denote the inverse of the IR regulated
propagators $\bar\Delta^{AB} = {\bar K}\Delta^{AB}$ with ${\bar K} =
1-K$.  For the aesthetic reason, we use the notation $\Phi^{*}_{A} =
\phi^{*}_{A}$ for the 1PI effective action.  
By adding a free part, we introduce the 1PI effective action $\Gamma$ as
\begin{eqnarray}
\Gamma = \frac{1}{2} \Phi^{A} \Delta^{-1}_{AB} \Phi^{B} 
+ \Phi_{A}^{*}R^{A}_{~~B} \Phi^{B}
+ \Gamma_{I}[\Phi,\Phi^{*}]\,.
\label{Gamma}
\end{eqnarray}
We also define the {\it total} 1PI effective action with regularized kinetic terms as
\begin{eqnarray}
\Gamma_{\rm tot} = \frac{1}{2} \Phi^{A} \bar\Delta^{-1}_{AB} \Phi^{B} 
+ \Phi^{*}_{A} R^{A}_{~~B}\Phi^{B}
+ \Gamma_{I}[\Phi,\Phi^{*}]\,.
\label{totalGamma}
\end{eqnarray}
Note the 1PI action $\Gamma$ and $\Gamma_{\rm tot}$ differ only in the
kinetic terms and the difference vanishes as the cutoff goes to zero.

Now we rewrite the QME in \bref{QME_S} in terms of the 1PI action.  
From \bref{Legendre} and (\ref{Legendre-1}), we find
\begin{eqnarray}
\frac{\partial^{r} S}{\partial\phi^A}K = \phi^{B}\Delta^{-1}_{BA} 
+ \phi_{B}^{*}R^{B}_{~~A}
+ \frac{\partial^{r} S_I}{\partial\phi^A}K 
= \frac{\pa^{r} \Gamma}{\pa\Phi^{A}}\,,  
~~~~~
\frac{\pa^{l} S_{I}}{\pa \phi^{*}_{A}} = 
\frac{\pa^{l} \Gamma_I}{\pa \Phi^{*}_{A}}\,.
\label{Legandre2}
\end{eqnarray}
Using \bref{Delta R relation}, \bref{Legandre2} and \bref{Gamma}, we
find\footnote{More detailed derivation will be found in
\cite{10.1093/ptep/ptz099}.}
\begin{eqnarray}
(S,S)_K = (\Gamma, \Gamma) \,.
\label{1st term of Sigma}
\end{eqnarray}
The antibracket on the r.h.s. is defined for arbitrary functionals of the classical fields $\Phi^{A}$
and their antifields $\Phi^{*}_{A}$ as
\begin{eqnarray}
(Z,W) = \frac{\pa^{r} Z}{\pa \Phi^{A}}\frac{\pa^{l} W}{\pa\Phi^{*}_{A}}
 - \frac{\pa^{r} Z}{\pa \Phi^{*}_{A}}\frac{\pa^{l} W}{\pa\Phi^{A}}\,.
\label{ab_Phi}
\end{eqnarray}
Note that the regulator function $K$ is absent in the above expression.

In rewriting the second term of QME, $\Delta_K S$, we note that only the
interaction action produces field dependent contributions. 
Using (\ref{Legandre2}), we obtain
\begin{eqnarray}
\Delta_{K} S_{I} = \frac{\pa^{r} }{\pa \phi^{A}}K 
\frac{\pa^{l} S_{I}}{\pa \phi^{*}_{A}} = \frac{\pa^{r} }{\pa \Phi^{B}}
\left(K \frac{\pa^{l} \Gamma_{I}}{\pa \Phi^{*}_{A}}\right)
\frac{\pa^{r}\Phi^{B}}{\pa \phi^{A}} 
= {\rm Tr}\left(K \Gamma^{(2)}_{I*}
\left[1 + \bar\Delta \Gamma^{(2)}_{I}\right]^{-1}\right)\,.
\label{Delta_S}
\end{eqnarray}
The last expression in \bref{Delta_S} is reached by using the relation
\begin{eqnarray}
\frac{\partial^{r}\Phi^A}{\partial\phi^B} = \left(\left[
1+\bar\Delta \Gamma^{(2)}_I \right]^{-1}\right)^{A}_{B}\,,
\label{Legendre1}
\end{eqnarray}
that is derived from (\ref{Legendre-1}).  Here we have used notations 
\begin{eqnarray}
\left(\Gamma^{(2)}_I\right)_{AB} = \frac{\pa^{l}\pa^{r}}{\pa\Phi^{A}\pa\Phi^{B}}
\Gamma_{I}\,,
\label{Gamma2}
\end{eqnarray}
as well as
\begin{eqnarray}
\left(\Gamma^{(2)}_{I*}\right)^{A}_{~~B} 
= \frac{\pa^{r} \pa^{l}}{\pa \Phi^{*}_{A}
\pa\Phi^{B}} \Gamma_{I}\,.
\label{Gamma2_star}
\end{eqnarray}

Finally, we find the modified Slavnov-Taylor (mST) identity as the
Legendre transform of the QME
\begin{eqnarray}
\Sigma = \frac{1}{2}(\Gamma,\Gamma) - {\rm Tr}\left(K \Gamma^{(2)}_{I*}
\left[1 + \bar\Delta \Gamma^{(2)}_{I}\right]^{-1}\right) =0 \,.
\label{QME_Gamma}
\end{eqnarray}

It is also worth pointing out that the second functional derivative of $\Gamma_{{\rm tot}}$ appeared in the second term of 
\bref{QME_Gamma} as
\begin{eqnarray}
1 + \bar\Delta \Gamma^{(2)}_{I} = \bar\Delta \Gamma_{{\rm tot}}^{(2)}\,.
\label{2nd derivative of Gatotal}
\end{eqnarray}
In \bref{QME_Gamma}, it is interesting to find $\Gamma$ in the first
term and $\Gamma_{{\rm tot}}$ in the second term.  Shortly we will find
a similar trace structure in the flow equation written for the 1PI action.

The measure operator $\Delta$ similar to \bref{AB} defined in terms of $\Phi^A$ and
$\Phi^*_A$ appears as the first-order part of eq. \bref{QME_Gamma}
\begin{eqnarray}
\Delta \Gamma = {\rm Tr}\left(K \Gamma_{I*}^{(2)}\right) \,.
\label{measure op for Gamma}
\end{eqnarray}

Here an important remark is in order. The Legendre transformation
(\ref{Legendre}) is not a canonical transformation from
$\{\phi^{A},\phi^{*}_{A}\}$ to $\{\Phi^{A},\Phi^{*}_{A}\}$.  Although
the antibracket \bref{ab_Phi} in terms of $\{\Phi^{A},\Phi^{*}_{A}\}$ is
convenient to write the relation \bref{1st term of Sigma}, one should
not mix up two canonical structures in $S$-world and $\Gamma$-world.

Using the flow equation for $S_{I}$ \cite{Polchinski:1983gv} 
\begin{eqnarray}
\dot{S_{I}} = \Lambda \pa_{\Lambda} S_I = - \frac{1}{2}\frac{\pa^{r} S_{I}}{\pa\phi^{A}}
\dot{\bar\Delta}^{AB}\frac{\pa^{l}S_{I}}{\pa\phi^{B}} + \frac{1}{2}
(-)^{A} \dot{\bar\Delta}^{AB} \frac{\pa^{l}\pa^{r}S_{I}}{\pa\phi^{B}\phi^{A}}\,
\label{flow_S}
\end{eqnarray}
and the Legendre transformation (\ref{Legendre}), \bref{Legendre-1}  and (\ref{Legendre1}),
we find that
\begin{eqnarray}
\dot{\Gamma_{I}} &=& \dot{S_{I}} + \frac{1}{2}\left(\phi-\Phi\right)^{A}
\left(\bar\Delta^{-1} \dot{\bar\Delta} \bar\Delta^{-1}\right)
\left(\phi-\Phi\right)^{B} 
\nn\\
&=& \dot{S_{I}} + \frac{1}{2}\frac{\pa^{r} S_{I}}{\pa\phi^{A}}
\dot{\bar\Delta}^{AB}\frac{\pa^{l}S_{I}}{\pa\phi^{B}}
= \frac{1}{2}
(-)^{A} \dot{\bar\Delta}^{AB} \frac{\pa^{l}\pa^{r}S_{I}}{\pa\phi^{B}\phi^{A}}\,.
\label{flow_to_Gamma}
\end{eqnarray}
Thus, we obtain the flow equation for $\Gammaeff$ \cite{Bonini:1992vh,Morris:1993qb,Wetterich:1992yh,Nicoll:1977hi} as
\begin{eqnarray}
\dot{\Gamma_{I}} = -\frac{1}{2}{\rm Str}\left(\dot{\bar\Delta}\bar\Delta^{-1}
\left[1 + \bar\Delta \Gamma^{(2)}_{I}\right]^{-1}\right) \,.
\label{flow_Gamma}
\end{eqnarray}

The expression of the Quantum Master Functional (QMF) $\Sigma$ in (\ref{QME_S}) 
and the flow equation (\ref{flow_S}) are combined to give 
\begin{eqnarray}
\dot{\Sigma} = - \frac{1}{2}\frac{\pa^{r} S_{I}}{\pa\phi^{A}}
\dot{\bar\Delta}^{AB}\frac{\pa^{l}\Sigma}{\pa\phi^{B}} + \frac{1}{2}
(-)^{A} \dot{\bar\Delta}^{AB} 
\frac{\pa^{l}\pa^{r}\Sigma}{\pa\phi^{B}\phi^{A}}\,.
\label{flow_Sigma}
\end{eqnarray}
That is, the QMF satisfies the linearized flow equation as a composite
operator \cite{Becchi:1996an} (See also \cite{Igarashi:2009tj}).  The
QME is stable along the RG flow once it holds at some cutoff scale.

In the next section, we consider QED with chiral invariant four-fermi
interactions and show that the QME/mST (\ref{QME_Gamma}) and the flow
equation (\ref{flow_to_Gamma}) can be simultaneously solved in a
perturbative expansion.

\section{1PI effective action in QED and the QME/mST}

\subsection{The classical effective action}

We consider 1PI effective action for QED with a massless Dirac fermion. 
The free part of the covariantly gauge fixed action contains kinetic
terms for the photon $A_{\mu}$, the Dirac field $\Psi, \bar\Psi$ and the
FP ghost fields $C$ and $\bar C$: the auxiliary field $B$ and the gauge
parameter $\xi$ are introduced accordingly.\footnote{We take the
gauge-fixed basis for antifields \cite{10.1093/ptep/ptz099}.}
In addition, here we also include antifields
$A^{*}_{\mu}$ and ${\bar C}^{*}$ as sources for the free BRST
transformations of $A_{\mu}$ and the antighost $\bar C$.  
\begin{eqnarray}
\Gamma_{0} = \int_{x}\biggl[
\frac{1}{2}\Bigl\{(\pa_{\mu}A_{\nu})^{2}- (\pa\cdot A)^{2}\Bigr\}
+ {\bar \Psi} i \Slash{\pa} \Psi 
+ \bigl(A^{*}_{\mu} -i \pa_{\mu} {\bar C}\bigr)\pa_{\mu}C + \frac{1}{2}\xi B^{2}
{+} \bigl({\bar C}^{*} {-} i \pa\cdot A\bigr)B\biggr]\,. 
\label{Gamma0}
\end{eqnarray}

Starting from $\Gamma_0$ in \bref{Gamma0}, we construct a
1PI effective action that satisfies the Classical
Master Equation
\begin{eqnarray}
(\Gamma_{{\rm cl}},~\Gamma_{{\rm cl}}) = 0
\label{CME}
\end{eqnarray}
up to ${\cal O}(e^2)$, $\Gamma_{{\rm cl}} = \Gamma_{0} + \Gamma_{1} +
\Gamma_2$.  The lower index is for the order of the gauge coupling.
The quantum part $\Gamma_q$ will be discussed later.
In order to solve eq. \bref{CME} we utilize the BRST cohomology argument
\cite{Fisch:1990uj,Henneaux:1991vp,Barnich:1995wt,Barnich:1995wm} that was applied earlier to Yang-Mills
theory in ERG \cite{10.1093/ptep/ptz099}.

From $(\Gamma_{1},\Gamma_{0}) =0$, we will uniquely determine
$\Gamma_1$, up to some normalization factors to be discussed later.  We
decompose $\Gamma_{1}$ into parts of definite antighost numbers.  The
table 1 is the list of various gradings of (anti)fields.  Looking for
local field combinations with the highest antighost number, mass
dimension four and of vanishing fermion and ghost numbers, we find the highest antighost number is one and
$\Gamma_{1}^{1} = \int_{x}[c_{1} \Psi^{*} \Psi C + c_{2} {\bar \Psi}^{*}
{\bar \Psi} C]$ with coefficients $c_1$ and $c_2$ to be determined
shortly.  The superscript of $\Gamma_{1}^{1}$ indicates the antighost number.  The only
candidate for $\Gamma_{1}^{0}$ is the minimal gauge interaction term
with the coupling $e$, $\Gamma_{1}^{0} = -e \int_{x}{\bar \Psi}
\Slash{A} \Psi$.  Now the requirement $(\Gamma_1^0+\Gamma_1^1,
\Gamma_0)=0$ fixes the coefficients $c_1$ and $c_2$ in $\Gamma_{1}^{1}$ as $c_{1} = -
c_{2} = - ie$.  In this manner, we find
\begin{eqnarray}
 \Gamma_{1} =\int_{x}
\biggl[ -e {\bar \Psi} \Slash{A} \Psi -i e \Psi^{*} \Psi C 
+i e {\bar \Psi}^{*} {\bar \Psi} C\biggr]\,.
\label{QED_Gamma01}
\end{eqnarray}
All contained in $\Gamma_{0} + \Gamma_{1}$ are marginal terms, 
and $\Lambda$ independent.

\begin{table}[ht]
\begin{center}
\begin{tabular}{ccccccc}
\hline
 & $\epsilon$ & gh \# & ag \# & pure gh \#  & dimension \\
\hline\hline
$A_\mu$ & 0 & 0 & 0 & 0 & $1$ \\
$C$ & 1 & 1 & 0 & 1 & $1$ \\
$\Psi,~{\bar \Psi}$ & 1 & 0 & 0 & 0 & $3/2$ \\
$\Psi^*,~{\bar \Psi}^*$ & 0 & -1 & 1 & 0 & $3/2$ \\
$\bar{C}$ & 1 & -1 & 1 & 0 & $1$  \\
$B$ & 0 & 0 & 1 & 1 & $2$\\
$A^*_{\mu}$ & 1 & -1 & 1 & 0 & $2$ \\
$\bar{C}^*$ & 0 & 0 & 0 & 0 & $2$\\
\hline
\end{tabular}
\end{center}
\caption{The various properties of the (anti)fields, namely, Grassmann parity, 
ghost number, antighost/antifield number, pure gh \# = gh \# + ag \#, and mass dimension. }
\label{table:ghostantighost}
\end{table}

We also include chiral invariant four-fermi interactions as irrelevant terms,
\begin{eqnarray}
\Gamma_{2,{\rm {\rm cl}}} &=& \int_{x} \biggl[\frac{G_{S}}{2\Lambda^{2}}\Bigl\{
\left({\bar\Psi}\Psi\right)\left({\bar\Psi}\Psi\right)
 - \left({\bar\Psi}\gamma_{5}\Psi\right)\left({\bar\Psi}
\gamma_{5}\Psi\right)\Bigr\}
\nn\\
&& 
+ \frac{G_{V}}{2\Lambda^{2}}\Bigl\{
\left({\bar\Psi}\gamma_{\mu}\Psi\right)\left({\bar\Psi}\gamma_{\mu}\Psi\right)
 + \left({\bar\Psi}\gamma_{5}\gamma_{\mu}\Psi\right)\left({\bar\Psi}
\gamma_{5}\gamma_{\mu}\Psi\right)\Bigr\}
\biggr]~\,.
\hspace{1cm}
\label{4fermi}
\end{eqnarray}
It is easy to confirm that $\Gamma_{{\rm cl}} = \Gamma_{0} + \Gamma_{1} +
\Gamma_{2}$ satisfies the Classical Master Equation in \bref{CME}.
In our perturbative expansion, we regard $G_{S},~G_{V}$ at the order of $e^{2}$.

The one-loop correction to the 1PI effective action is given 
as the closed-form solution to (\ref{flow_Gamma}):
\begin{eqnarray}
&& 
\Gamma_{{\rm q}} = \frac{1}{2} {\rm Str}\log \left(\bar\Delta^{-1} 
+ \Gamma_{I, {\rm cl}}^{(2)}\right)\,,
\label{Str_log_formula}
\end{eqnarray}
where $\Gamma_{I, {\rm cl}}^{(2)}$ is the classical part of
\bref{Gamma2}, the second functional derivative of $\Gamma_1 +
\Gamma_2$~.  ${\bar \Delta}$ in \bref{Str_log_formula} are 
the IR-regularized propagators,
\begin{eqnarray}
\bar\Delta_{\mu\nu} 
= (P^{T}_{\mu\nu} + \xi P^{L}_{\mu\nu}){\mathbf {\bar\Delta}}\,, 
~~~~~~
\bar\Delta_{\alpha\hat\beta} = 
(i \Slash{\pa})_{\alpha\hat\beta} {\mathbf {\bar\Delta}} 
\,,
\label{IRreg_prop}
\end{eqnarray}
for the gauge and Dirac fields respectively.  Here, ${\mathbf
{\bar\Delta}} = (1-K)/(-\pa^{2})$, $P^{T}_{\mu\nu}$ and $P^{L}_{\mu\nu}$
are the transverse and longitudinal projection operators.  The lowest
order quantum correction is simply
\begin{eqnarray}
\Gamma_{1,{\rm q}} = \frac{1}{2}{\rm Str}\bigl({\bar \Delta} \Gamma^{(2)}_1\bigr)\,.
\label{Gamma1q}
\end{eqnarray}
The r.h.s. of \bref{Gamma1q} is evaluated with first two vertices in \bref{vertices} of Appendix A
and turned out to be zero.  The perturbative expansion of \bref{Str_log_formula} starts from ${\cal O}(e^2)$ term.

We expand the QMF according to the order of couplings as
\begin{eqnarray}
\Sigma = \Sigma_{0} + \Sigma_{1} + \Sigma_{2} + \Sigma_{3} + \cdots\,,
\label{Sigma_expan}
\end{eqnarray}
and we find 
$\Sigma_{0} = (\Gamma_{0},\Gamma_{0})/2 - \Delta \Gamma_{0} =0$ and 
$\Sigma_{1} =(\Gamma_{1},\Gamma_{0}) - \Delta \Gamma_{1} =0$ with $\Delta$ defined in eq. \bref{measure op for Gamma}.  

In the following two subsections, we evaluate $\Sigma_{2,3}$, higher order terms in
\bref{Sigma_expan}, after obtaining quantum corrections,
$\Gamma_{2,{\rm q}}$ and $\Gamma_{3,{\rm q}}$.

\subsection{Second order in gauge coupling}

Expanding eq. (\ref{Str_log_formula}) to the orders of $e^{2}$ and $G_{S,V}$, we
obtain a quantum part of the action
\begin{eqnarray}
\Gamma_{2,{\rm q}} = 
\frac{1}{2}{\rm Str}\Bigl(\bar\Delta \Gamma_{2,{\rm cl}}^{(2)}\Bigr) +
{\rm Str}\Bigl(-\frac{1}{4} \bar\Delta\Gamma_{1}^{(2)} \bar\Delta\Gamma_{1}^{(2)} \Bigr) 
\label{Gamma_2q}\,,
\end{eqnarray}
which has gauge and fermion fields contributions.  We write them separately as 
\begin{eqnarray}
\Gamma_{2,{\rm q}}^{AA} =  \frac{1}{2} \bar\Delta_{\al\hat\al}
\tau_{\hat\al\beta}^{(-\Slash{A})}
\bar\Delta_{\beta\hat\beta}\tau_{\hat\beta\al}^{(-\Slash{A})}
= \frac{1}{2} e^{2} 
\Bigl[(i \Slash{\pa}){\mathbf {\bar\Delta}} \Slash{A} (i \Slash{\pa})
{\mathbf {\bar\Delta}}  \Slash{A} \Bigr]
\label{AA}
\end{eqnarray}
and
\begin{eqnarray}
\Gamma_{2,{\rm q}}^{\bar\Psi\Psi} &=& - \bar\Delta_{\al\hat\beta}
\tau^{(\bar\Psi\Psi)}_{\hat\beta\al}
+ \bar\Delta_{\al\hat\al}
\tau_{\hat\al\mu}^{(-\gamma\Psi)} \bar\Delta_{\mu\nu} 
\tau_{\nu\al}^{(- \bar\Psi\gamma)}
\nn\\
&=& 
- e^{2} \Bigl[\bar\Delta_{\mu\nu} \bar\Psi\gamma_{\nu} 
(i \Slash{\pa}) {\mathbf {\bar\Delta}} \gamma_{\mu} \Psi\Bigr]~.
\label{barPsi_Psi}
\end{eqnarray}
Here the quantities $\tau$ are the vertices obtained from
$\Gamma^{(2)}$ listed in Appendix A.  Note that the
four-fermi interactions give vanishing contribution in
eq. (\ref{barPsi_Psi}): in momentum space, it becomes
\begin{eqnarray}
\bar\Delta_{\al\hat\beta}\tau^{(\bar\Psi\Psi)}_{\hat\beta\al} \rightarrow 
\frac{2}{\Lambda^{2}} (G_{S}-4G_{V}) 
\int_{p,q}
\bar\Psi(-p)\gamma_{\mu}\Psi(p)
\frac{1-K(q)}{q^{2}}q_{\mu} =0\,.
\label{4_fermi_barPsi_Psi}
\end{eqnarray}
The integral over $q$ vanishes due to the Lorentz covariance.
\footnote{Strictly speaking, this happens due to a cancellation of
divergent contributions, because $\int_{q} 1/q^{2}$ is UV
divergent.  We may regularize the $q^{2}$ integral
$\int_{0}^{\infty} dq^{2}$ as $\int_{0}^{\Lambda_{0}^{2}} dq^{2}$ with a
UV cutoff $\Lambda_{0}$ or we may instead use the dimensional
regularization.}

Having constructed $\Gamma_{2,q}$, we may calculate the QMF at ${\cal O}(e^{2})$,
\begin{eqnarray}
\Sigma_{2} = (\Gamma_{0}, \Gamma_{2,{\rm q}})   
+ 
\Bigl[K \Gamma_{1*}^{(2)} \bar\Delta\Gamma_{1}^{(2)} \Bigr]\,,
\label{Sigma_2} 
\end{eqnarray}
where $\Gamma_{1*}^{(2)}$ is ${\cal O}(e)$ part of $\Gamma_{I*}^{(2)}$.
It turned out that both terms of $\Sigma_{2}$ are proportional to $A_\mu C$, the gauge
field multiplied by the ghost.
The second term in (\ref{Sigma_2}) becomes 
\begin{eqnarray}
\Sigma_{2}|_{K} &=&
\Bigl[K \Gamma_{1*}^{(2)} \bar\Delta\Gamma_{1}^{(2)} \Bigr]
= 
\Bigl[ K\tau^{C}_{*\al\beta}\bar\Delta_{\beta\hat\beta} 
\tau_{\hat\beta\al}^{(-\Slash{A})}
+ K \tau^C_{*{\hat\al}{\hat\beta}}
{\bar \Delta}^t_{{\hat\beta}\al} \tau_{\al{\hat\al}}^{(\Slash{A}^t)} 
\Bigr]
\nn\\
&=&  e^{2}
\Bigl[K  C \Slash{\pa}{\mathbf {\bar\Delta}}  \Slash{A} 
- \Slash{A}\Slash{\pa}{\mathbf {\bar\Delta}} C K  
\Bigr]
= 8 e^{2} \Bigl[ K C \pa_{\mu} \bar{\mathbf \Delta} A_{\mu}
\Bigr] \,.
\label{SigmaACK} 
\end{eqnarray}
On the other hand, as shown in Appendix B, the first term of eq. 
(\ref{Sigma_2}) becomes
\begin{eqnarray}
\Sigma_{2}|_{(\Gamma_{0}, \Gamma_{2,{\rm q}}) } &=& - \frac{e^{2}}{2} {\rm tr}\bigl(
\gamma_{\mu}\gamma_{\nu}\gamma_{\rho}\gamma_{\sigma}\bigr)
\Bigl[\pa_{\mu} \bar{\mathbf \Delta} A_{\nu} \pa_{\rho} \bar{\mathbf \Delta} \pa_{\sigma} C 
+ \pa_{\mu} \bar{\mathbf \Delta} \pa_{\nu} C \pa_{\rho} \bar{\mathbf \Delta} A_{\sigma}  \Bigr]
\nn\\
&=& 8 e^{2} \Bigl[(1-K)C\pa_{\nu} \bar{\mathbf \Delta} A_{\nu}
\Bigr] \,.
\label{Sigma_AC}
\end{eqnarray}
Therefore, 
\begin{eqnarray}
\Sigma_{2} &=& \Sigma_{2}|_{(\Gamma_{0}, \Gamma_{2}) } 
+ \Sigma_{2}|_{K} = 8 e^{2} \Bigl[C \pa_{\mu}\bar
{\mathbf \Delta} A_{\mu} 
\Bigr] \nn\\
 &=& - 8i e^{2}\int_{p,q} C(p) \frac{[1 - K(q)]}{q^{2}}q_{\mu}
A_{\mu}(-p) =0\,.
\label{Sigma20}
\end{eqnarray}
We have shown that the QME and the flow equation are consistently solved
at ${\cal O}(e^{2})$ and ${\cal O}(G_{S,V})$.

$\Gamma_{2,{\rm q}}^{AA}$ given in eq. \bref{AA} may be written as
\begin{eqnarray}
\Gamma_{2,{\rm q}}^{AA} = \frac{1}{2} e^{2} \int_{p} A_{\mu}(-p) \left[
P^{T}{\cal T}(p) + P^{L}{\cal L}(p)\right]A_{\nu}(p)\,.
\label{AA1}
\end{eqnarray}
Its longitudinal part ${\cal L}$ 
\begin{eqnarray}
{\cal L}(p) = -8 \int_{q} K(p+q)[1-K(q)]\frac{(p\cdot q)}{p^{2} q^{2}}\,
\label{longitudinal}
\end{eqnarray}
is necessary to satisfy the QME at ${\cal O}(e^2)$, $\Sigma_{2} =0$.
Once we put the IR cutoff $K=0$ by sending $\Lambda \rightarrow 0$, we recover the standard WT relation ${\cal L}(p) =0$.

\subsection{Third order in gauge coupling}
 
Expanding (\ref{Str_log_formula}) to  ${\cal O}(e^{3})$ and ${\cal O}(e G_{S,V})$
, we obtain
\begin{eqnarray}
\Gamma_{3,{\rm q}} = 
\frac{1}{6}
{\rm Str}\Bigl( \bar\Delta \Gamma_{1}^{(2)}
\bar\Delta\Gamma_{1}^{(2)}\bar\Delta \Gamma_{1}^{(2)} 
\Bigr)
-\frac{1}{4} {\rm Str}
\Bigl(\bar\Delta\Gamma_{1}^{(2)}
\bar\Delta\Gamma_{2,{\rm cl}}^{(2)} \Bigr)
\,.
\label{Gamma3}
\end{eqnarray}
This gives quantum corrections to $\bar\Psi \Slash{A} \Psi$ vertex.  Two 
terms in \bref{Gamma3} are proportional to $e^{3}$ and $eG_{S,V}$ respectively,
\begin{eqnarray}
\Gamma_{3,{\rm q}} = \Gamma_{3,{\rm q}}|_{e^{3}} + \Gamma_{3,{\rm q}}|_{eG}\,,
\label{Gamma3-2}
\end{eqnarray}
where
\begin{eqnarray}
\Gamma_{3,{\rm q}}|_{e^{3}}
&=&  \frac{1}{3}\Bigl(- \bar\Delta_{\al\hat\al} 
\tau_{\hat\al\mu}^{(-\gamma\Psi)} \bar\Delta_{\mu\nu} 
\tau_{\nu\beta}^{(- \bar\Psi\gamma)}\bar\Delta_{\beta\hat\beta} 
\tau_{\hat\beta\al}^{(-\Slash{A})} 
\nn\\
&& 
- \bar\Delta_{\al\hat\al}
\tau_{\hat\al\beta}^{(-\Slash{A})} \bar\Delta_{\beta\hat\beta}
\tau_{\hat\beta\mu}^{(-\gamma\Psi)}\bar\Delta_{\mu\nu} 
\tau_{\nu\al}^{(- \bar\Psi\gamma)} 
+ \bar\Delta_{\mu\nu} \tau_{\nu\al}^{(- \bar\Psi\gamma)}
\bar\Delta_{\al\hat\al}\tau_{\hat\al\beta}^{(-\Slash{A})} 
\bar\Delta_{\beta\hat\beta} \tau_{\hat\beta\mu}^{(-\gamma\Psi)}\Bigr) 
\nn\\
&=& -e^{3}\Bigl[ \bar\Delta_{\mu\nu} 
\bar\Psi \gamma_{\nu} (i \Slash{\pa}){\mathbf {\bar\Delta}} \Slash{A} (i \Slash{\pa})
{\mathbf {\bar\Delta}} \gamma_{\mu} \Psi\Bigr]\,,  \nn\\
\Gamma_{3,{\rm q}}|_{eG} &=& \bar\Delta_{\al\hat\al}
\tau^{(-\Slash{A})}_{\hat\al\beta}\bar\Delta_{\beta\hat\beta}
\tau^{\bar\Psi\Psi}_{\hat\beta\al}   \nn\\
&=& 2 e \frac{G_{S}}{\Lambda^{2}}\Bigl[\bar\Psi (i \Slash{\pa})
{\mathbf {\bar\Delta}} \Slash{A} (i \Slash{\pa}){\mathbf {\bar\Delta}} \Psi\Bigr]
+ 2e \frac{G_{V}}{\Lambda^{2}} \Bigl[\bar\Psi \gamma_{\mu} 
(i \Slash{\pa}){\mathbf {\bar\Delta}} \Slash{A} 
(i \Slash{\pa}){\mathbf {\bar\Delta}}
\gamma_{\mu} \Psi \Bigr]\,,  \nn\\
&& -4 e \frac{G_{V}}{\Lambda^{2}}
\Bigl[i\pa_{\rho} \bar{\mathbf \Delta} A_{\rho} i \pa_{\mu}\bar{\mathbf \Delta}
- i\pa_{\rho} \bar{\mathbf \Delta} A_{\mu} i \pa_{\rho}\bar{\mathbf \Delta}
+ i\pa_{\mu} \bar{\mathbf \Delta} A_{\rho} i \pa_{\rho}\bar{\mathbf \Delta}
\Bigr]\Bigl(\bar\Psi \gamma_{\mu} \Psi\Bigr)\,.
\label{Gamma3_e3eG}
\end{eqnarray}

Now we may calculate QME at $ {\cal O}(e^{3})$ and ${\cal O}(e G_{S,V})$ as
\begin{eqnarray}
\Sigma_{3} = (\Gamma_{3,{\rm q}}, \Gamma_{0}) + (\Gamma_{1}, \Gamma_{2,{\rm q}}) 
+ 
\Bigl[K \Gamma_{1*}^{(2)} \bar\Delta \Gamma_{2,{\rm cl}}^{(2)}\Bigr]
- 
\Bigl[
K \Gamma_{1*}^{(2)} \bar\Delta\Gamma_{1}^{(2)}\bar\Delta\Gamma_{1}^{(2)} 
\Bigr]\,.
\label{Sigma_3} 
\end{eqnarray}
All the terms in $\Sigma_3$ are proportional to $\bar\Psi\Psi C$ with coefficients of ${\cal O}(e^3)$ or ${\cal O}(eG)$: $\Sigma_{3} = 
\Sigma_{3,e^{3}}+ \Sigma_{3,eG}$.  
There are three ${\cal O}(e^3)$ terms 
\begin{eqnarray}
\Sigma_{3,e^{3}}
= \Sigma_{3,e^{3}}
|_{K}
 + \Sigma_{3,e^{3}}
|_{(\Gamma_{0},\Gamma_{3,{\rm q}})} 
+ \Sigma_{3,e^{3}}
|_{(\Gamma_{1},\Gamma_{2,{\rm q}})} \,
\label{Sigma3e3_orig}
\end{eqnarray}
where
\begin{eqnarray}
\Sigma_{3,e^{3}}|_{K} &=& - 
\Bigl[
K \Gamma_{1*}^{(2)} \bar\Delta\Gamma_{1}^{(2)}\bar\Delta\Gamma_{1}^{(2)} \Bigr]
\nn\\
&=& -  
\Bigl[K \tau^{C}_{*\al\beta} \bar\Delta_{\beta\hat\al} 
\tau_{\hat\al\mu}^{(-\gamma\Psi)}\bar\Delta_{\mu\nu} 
\tau_{\nu\al}^{(-\bar\Psi\gamma)}  + K 
\tau^{C}_{*\hat\al\hat\beta} \bar\Delta_{\hat\beta\al} 
\tau_{\al\mu}^{(\bar\Psi\gamma)}\bar\Delta_{\mu\nu} 
\tau_{\nu\hat\al}^{(\gamma\Psi)}\Bigr] 
\nn\\
&=& e^{3} \Bigl[\bar\Psi\gamma_{\nu} KC \Slash{\pa}{\mathbf {\bar\Delta}}
\gamma_{\mu} \Psi \bar\Delta_{\mu\nu}\Bigr] -  e^{3} \Bigl[\bar\Psi\gamma_{\mu} 
\Slash{\pa}{\mathbf {\bar\Delta}} C K  \gamma_{\nu} \Psi\bar\Delta_{\mu\nu}\Bigr] \,,
\label{Sigma_bPPCK}\\
\Sigma_{3,e^{3}}|_{(\Gamma_{1},\Gamma_{2,{\rm q}})} &=& - \frac{\pa\Gamma_{1}}{\pa \Psi^{*}}
\frac{\pa^{l}\Gamma_{2,q}}{\pa \Psi} - \frac{\pa\Gamma_{1}}{\pa \bar\Psi^{*}}
\frac{\pa^{l}\Gamma_{2,q}}{\pa \bar\Psi} 
\nn\\
&=& 
e^{3} \Bigl[\bar\Psi\gamma_{\nu}  \Slash{\pa}
{\mathbf {\bar\Delta}}
C\gamma_{\mu} \Psi \bar\Delta_{\mu\nu}\Bigr]
- e^{3}\Bigl[ \bar\Psi  \gamma_{\nu}C  \Slash{\pa}
{\mathbf {\bar\Delta}}
\gamma_{\mu} \Psi \bar\Delta_{\mu\nu}\Bigr]\,,
\label{S3_bPPC}\\
%
\Sigma_{3,e^{3}}|_{(\Gamma_{0},\Gamma_{3,{\rm q}})} &=& - e^{3} \Bigl[
\bar\Psi \gamma_{\nu} \Slash{\pa}{\mathbf {\bar\Delta}}
\Slash{\pa}C \Slash{\pa}{\mathbf {\bar\Delta}}
\gamma_{\mu} \Psi\bar\Delta_{\mu\nu}\Bigr]
\nn\\
 &=&  e^{3} 
\Bigl[\bar\Psi \gamma_{\nu} (1-K) C \Slash{\pa}{\mathbf {\bar\Delta}}
\gamma_{\mu} \Psi\bar\Delta_{\mu\nu}\Bigr] - e^{3} \Bigl[
\bar\Psi \gamma_{\nu} \Slash{\pa}{\mathbf {\bar\Delta}}
C (1-K)
\gamma_{\mu} \Psi \bar\Delta_{\mu\nu} \Bigr]\,.
\hspace{1cm}
\label{S3s0}
\end{eqnarray}
The above results lead to
\begin{eqnarray}
&& 
\Sigma_{3,e^{3}} = 0\,.
\label{Sigma3e3}
\end{eqnarray}
As for ${\cal O}(eG)$ terms, we find two contributions
\begin{eqnarray}
\Sigma_{3,eG} = \Sigma_{3,eG}|_{K}
 + \Sigma_{3,eG}|_{(\Gamma_{0},\Gamma_{3,q})} 
\,.
\label{Sigma3eG_orig}
\end{eqnarray}
We may calculate them as 
\begin{eqnarray}
&& \Sigma_{3,eG}|_{K} =
\Bigl[ K \Gamma_{1*}^{(2)} \bar\Delta\Gamma_{2,{\rm cl}}^{(2)} \Bigr]
=  
\Bigl[
K \tau^{C}_{*\al\beta}\bar\Delta_{\beta\hat\al}
\tau^{(\bar\Psi\Psi)}_{\hat\al\al} + K \tau^{C}_{*\hat\al\hat\beta}
\bar\Delta^{T}_{\hat\beta\al}\left(\tau^{(\bar\Psi\Psi)}\right)^{T}_{\al\hat\al}
\Bigr]
\label{Sigma3GPPCK}\\
&&~~~= \frac{2e (G_S-2 G_V)}{\Lambda^2}
\Bigl[ \bar\Psi \Bigl(KC \Slash{\pa}{\mathbf {\bar\Delta}}
- \Slash{\pa}{\mathbf {\bar\Delta}}
CK\Bigr)\Psi \Bigr]
\nn\\
&&~~~~~~~~~~-\frac{4e G_{V}}{\Lambda^{2}}
\Bigl[ \Bigl(KC \pa_{\mu}\bar{\mathbf \Delta} - \pa_{\mu}\bar{\mathbf \Delta} CK\Bigr)
(\bar\Psi \gamma_{\mu}\Psi) \Bigr] \,,  \nn
\end{eqnarray}
and 
\begin{eqnarray}
\Sigma_{3,eG}|_{(\Gamma_{0},\Gamma_{3,{\rm q}})}  &=& 
\frac{2e (G_S-2 G_V)}{\Lambda^2}
\Bigl[ \bar\Psi\Bigl((1-K)C \Slash{\pa}{\mathbf {\bar\Delta}} - \Slash{\pa}{\mathbf {\bar\Delta}}
C(1-K)\Bigr)\Psi \Bigr] \nn\\
&& - \frac{4e G_{V}}{\Lambda^{2}}
\Bigl[ \Bigl((1-K)C\pa_{\mu}\bar{\mathbf\Delta} - \pa_{\mu}\bar{\mathbf\Delta} C(1-K)\Bigr)
(\bar\Psi\gamma_{\mu}\Psi) \Bigr] \,.
\label{Sigma3GbPPC}
\end{eqnarray}
Eqs. \bref{Sigma3GPPCK} and \bref{Sigma3GbPPC} sum up to give a vanishing result,
\begin{eqnarray}
\Sigma_{3,eG} = 
\frac{2e (G_S-2 G_V)}{\Lambda^2}
\Bigl[
\bar\Psi\bigl(C \Slash{\pa}{\mathbf {\bar\Delta}} - \Slash{\pa}{\mathbf {\bar\Delta}} C\bigr)
\Psi
\Bigr]
- \frac{4e G_{V}}{\Lambda^{2}}
\Bigl[
\bigl(C\pa_{\mu}\bar{\mathbf\Delta} - \pa_{\mu}\bar{\mathbf\Delta} C\bigr)
(\bar\Psi\gamma_{\mu}\Psi)
\Bigr]=0
\,.
\label{Sigma3eG}
\end{eqnarray}

From \bref{Sigma3e3} and \bref{Sigma3eG}, we finally obtain the result
\begin{eqnarray}
\Sigma_{3} =0 \,.
\label{Sigma3_total}
\end{eqnarray}
We have confirmed that the QME/mST can be solved consistently with the
flow equation up to the orders of $e^{3}$ and $eG_{S,V}$.  We have seen that the four-fermi
interactions generate quantum corrections to $\bar\Psi \Slash{A} \Psi$
vertex function and, in $\Sigma_{3}$, their free BRST transformation and the
measure factor cancel each other.

\section{Wavefunction renormalization constants and $\beta$ functions}

In order to take account of $\Lambda$ evolution, we introduce
renormalization constants for fields and couplings.  The corresponding
$Z$ factors are defined as
\begin{eqnarray}
&& 
A_{\mu} \rightarrow Z_{3}^{1/2} A_{\mu}, \qquad \Psi \rightarrow Z_{2}^{1/2}
\Psi,
\qquad \bar\Psi \rightarrow Z_{2}^{1/2}\bar\Psi
, \qquad 
C \rightarrow Z_{3}^{1/2} C, 
\nn\\
&& 
 {\bar C} \rightarrow Z_{3}^{-1/2} {\bar C} 
, \qquad B \rightarrow Z_{3}^{-1/2} B, \qquad 
A^{*}_{\mu} \rightarrow Z_{3}^{-1/2} A^{*}_{\mu}
\nn
\\
&& 
\Psi^{*} \rightarrow Z_{2}^{-1/2}
\Psi^{*}, \qquad \bar\Psi^{*} \rightarrow Z_{2}^{-1/2}
\Psi^{*}, \qquad {\bar C}^{*} \rightarrow Z_{3}^{1/2}{\bar C}^{*}\,.
\label{Z_factor}
\end{eqnarray}
For the gauge coupling, four-fermi couplings and gauge parameter, we set
$e \rightarrow e_{\Lambda}=Z_{e} e, G_{S,V} \rightarrow Z_{S,V}G_{S,V}$
and $\xi \rightarrow Z_{3}\xi$. There is some ambiguity in introducing
wavefunction renormalization factors. Here, we require that the $Z$
factors in eq. \bref{Z_factor} to be wavefunction rescalings due to
canonical transformations so that the fields and antifields are rescaled
in opposite directions \cite{10.1093/ptep/ptz099}.

The 1PI effective action is expressed as
\begin{eqnarray}
&& 
\Gamma_{0} = \int_{x}\biggl[\frac{Z_3}{2} A_{\mu} 
\bigl(-\pa^{2} P^{T}_{\mu\nu}\bigr)A_{\nu}
+ Z_{2}{\bar \Psi} i \Slash{\pa} \Psi + 
\bigl(A^{*}_{\mu} -i \pa_{\mu} {\bar C}\bigr)\pa_{\mu}C 
+ \frac{\xi}{2} B^{2}
+ \bigl({\bar C}^{*} + i \pa\cdot A\bigr)B\biggr]~, 
\nn\\
&& 
\Gamma_{1} = \int_{x}\biggl[-e(Z_{e}Z_{3}^{1/2}Z_{2}) {\bar \Psi} 
\Slash{A} \Psi -i e
(Z_{e}Z_{3}^{1/2}) \Psi^{*} \Psi C 
+i e (Z_{e}Z_{3}^{1/2}) {\bar \Psi}^{*} {\bar \Psi} C\biggr]~,
\nn\\
&& 
\Gamma_{2} = \int_{x}\biggl[(Z_{S}^{1/2}Z_{2}^{2})
\frac{G_{S}}{2\Lambda^{2}}\Bigl\{
\left({\bar\Psi}\Psi\right)\left({\bar\Psi}\Psi\right)
 - \left({\bar\Psi}\gamma_{5}\Psi\right)\left({\bar\Psi}
\gamma_{5}\Psi\right)\Bigr\}
\nn\\
&& 
~~~~~+ (Z_{V}^{1/2}Z_{2}^{2})^{-1}\frac{G_{V}}{2\Lambda^{2}}\Bigl\{
\left({\bar\Psi}\gamma_{\mu}\Psi\right)\left({\bar\Psi}\gamma_{\mu}\Psi\right)
 + \left({\bar\Psi}\gamma_{5}\gamma_{\mu}\Psi\right)\left({\bar\Psi}
\gamma_{5}\gamma_{\mu}\Psi\right)\Bigr\}
\biggr]\,.
\label{QED_Gamma012_Z}
 \end{eqnarray}

At one loop level anomalous dimensions 
of the photon and Dirac fields are expressed as 
\begin{eqnarray}
Z_{2,3} = 1 - \eta_{\Psi, A} \log (\Lambda/\mu)\,.
\label{anomalous.dim}
\end{eqnarray}
For the gauge coupling, its beta function is expressed as 
$\beta_{e} = \eta_{e} e$ where
\begin{eqnarray}
Z_{e} = 1 + \eta_{e} \log (\Lambda/\mu)\,.
\label{Ze}
\end{eqnarray}

We first compute $\eta_{A,\Psi,e}$ in the absence of the four-fermi interactions, 
$G_{S} = G_{V} =0$.  For the photon two-point functions,
it follows from (\ref{AA}) in momentum space 
\begin{eqnarray}
{\dot \Gamma}_{2}^{AA} &=& e_{\Lambda}^{2} \int_{p,q}{\rm Tr}\biggl\{
\frac{{\dot {\bar K}}(q) {\bar K}(p+q)}{q^{2}(p+q)^{2}}
\bigl[\Slash{q}\Slash{A}(-p)
(\Slash{p} + \Slash{q})\Slash{A}(p)\bigr]
\biggr\} 
\nn\\
&=& \frac{e_{\Lambda}^{2}}{2} \int_{p}A_{\mu}(-p) {\dot {\cal A}}_{\mu\nu}(p) 
A_{\nu}(p)\,.
\label{gammaAA2dot}
\end{eqnarray}
We expand ${\cal A}_{\mu\nu}$  in external momentum up to $p^{2}$. 
${\dot {\cal A}}_{\mu\nu}(0)$ gives a photon mass term
\begin{eqnarray}
M_{A}^{2} = \frac{e_{\Lambda}^{2}}{4\pi^{2}} \Lambda^{2} \int_{0}^{\infty}
du~u{\bar K}'(u){\bar K}(u)\,,
\label{mass}
\end{eqnarray}
while ${\cal O}(p^{2})\propto P^{T}_{\mu\nu} p^{2}$ part yields
\begin{eqnarray}
\eta_{A} = - \frac{e_{\Lambda}^{2}}{6\pi^{2}}
\int_{0}^{\infty}du \Bigl\{\bigl[(u {\bar K(u)}^{\prime})^{2}\bigr]^{\prime}-
({\bar K}^{2}(u))^{\prime}\Bigr\} = \frac{e_{\Lambda}^{2}}{6\pi^{2}}~\,.
\label{etaA}
\end{eqnarray}

In momentum space  $\Lambda \pa_{\Lambda} = \pa_{t}$ derivative of the
fermion two-point function (\ref{barPsi_Psi}) takes the 
form 
\begin{eqnarray}
{\dot \Gamma}^{\bar\Psi\Psi}_{2} = 
-e_{\Lambda}^{2} \int_{p} {\bar\Psi}(-p){\dot \Gamma}_{2}^{\bar\Psi\Psi}(p) \Psi(p)\,,
\label{GammaPsi_2}
\end{eqnarray}
where
\begin{eqnarray}
{\dot \Gamma}^{\bar\Psi\Psi}_{2}(p) = \int_{q}
\gamma_{\nu} (\Slash{p} +\Slash{q} )\gamma_{\mu} 
\frac{\dot{\bar K}(q){\bar K}(p+q) + 
{\bar K}(q)\dot{\bar K}(p+q)}{q^{2}(p+q)^{2}}
\Bigl(\delta_{\mu\nu} + 
(\xi -1)\frac{q_{\mu}q_{\nu}}{q^{2}}\Bigr)\,.
\label{gammapp2dot}
\end{eqnarray}
It gives
\begin{eqnarray}
\frac{\Slash{p}}{p^{2}}{\dot \Gamma}^{\bar\Psi\Psi}_{2}(p)|_{p^{2}=0} 
= \frac{\xi}{8\pi^{2}} \int_{0}^{\infty} du [{\bar K}^{2}(u)]' = 
\frac{\xi}{8\pi^{2}}  \,.
\label{gammapp2dot1}
\end{eqnarray}
Therefore, the anomalous dimension for the Dirac fields is
\begin{eqnarray}
\eta_{\Psi} = \frac{e_{\Lambda}^{2}\xi}{8\pi^{2}}  \,.
\label{etapsi}
\end{eqnarray}

For the gauge interaction vertex, we have 
\begin{eqnarray}
{\dot \Gamma}_{3}^{\bar\Psi A \Psi}  
= -e_{\Lambda}^{3} \int_{p,q,r} {\dot \Gamma}_{3,\rho}^{\bar\Psi A \Psi}(p,q,r)
\bar\Psi(p) A_{\rho}(q)\Psi(r)\delta(p+q+r)\,,
\label{PPA50}
\end{eqnarray}
where 
\begin{eqnarray}
{\dot \Gamma}_{3,\rho}^{\bar\Psi A \Psi}(0,0,0) &=& \frac{\pa}{\pa t}
\int_{q}\frac{{\bar K}^{3}(q)}{q^{6}} \gamma_{\nu} \Slash{q}\gamma_{\rho} 
\Slash{q}\gamma_{\mu}\Bigl(\delta_{\mu\nu} + 
(\xi -1)\frac{q_{\mu}q_{\nu}}{q^{2}}\Bigr)
\nn\\
&=& -
\frac{2 \xi \gamma_{\rho}}{(4\pi)^{2}} \int_{0}^{\infty} du ~({\bar K}^{3}(u))^{\prime}
= -\frac{\xi \gamma_\rho}{8 \pi^2}\,.
\label{PPA51}
\end{eqnarray}
Therefore,  
\begin{eqnarray}
{\dot \Gamma_{3}}^{\bar\Psi A \Psi}  
= 
\frac{e_{\Lambda}^{3}\xi}{8\pi^{2}} \bar\Psi \Slash{A} \Psi  \,.
\label{PPA52}
\end{eqnarray}

The flow equation for the gauge interaction vertex takes the form
\begin{eqnarray}
e_{\Lambda} \left(\frac{1}{2}\eta_{A} + \eta_{\Psi} - \eta_{e}\right)
=  \frac{e_{\Lambda}^{3}\xi}{8\pi^{2}}\,,
\label{flow_e}
\end{eqnarray}
which leads to 
\begin{eqnarray}
\eta_{e} = \frac{1}{2}\eta_{A} = \frac{e_{\Lambda}^{2}}{12\pi^{2}}\,.
\label{eta_e}
\end{eqnarray}
This is equivalent to the well-known Ward identity
\begin{eqnarray}
Z_{1} = Z_{3}^{1/2} Z_{e} Z_{2} = Z_{2}\,,
\label{WT_rel}
\end{eqnarray}
in our one-loop computation.

The beta function for the gauge coupling is given by
\begin{eqnarray}
{\dot e_{\Lambda}} = \beta_{e} = \eta_{e} e_{\Lambda} 
= \frac{e_{\Lambda}^{3}}{12\pi^{2}}\,.
\label{beta_e}
\end{eqnarray}

We stress that the anomalous dimensions $\eta_{A,\Psi}$ and the beta function 
$\beta_{e}$ are universal, being independent of the choice of cutoff function 
$K$. They are the same as those obtained by using gauge invariant 
regularization as the dimensional regularization in 
the standard perturbation theory. 

We now include the four-fermi interactions, $G_{S}$ and $G_V$. The
presence of these couplings leave the anomalous dimensions
$\eta_{A,\Psi}$ unchanged, while there arises an additional contribution
in the r.h.s. of the flow eq. (\ref{flow_e}). Instead of (\ref{eta_e}),
we have
\begin{eqnarray}
\eta_{e} = \frac{1}{2}\eta_{A} - \frac{1}{4\pi^{2}}(G_{S} -4 G_{V})
\int_{0}^{\infty}du~ u {\bar K(u)}{\bar K(u)}'\,.
\label{mdWT}
\end{eqnarray}
 The coefficient of $G_{S} -4 G_{V}$ depends on choice of the 
cutoff function, but is related to the photon mass term. Therefore, we obtain 
\begin{eqnarray}
\eta_{e} = \frac{1}{2}\eta_{A}  - 
{\bar M_{A}}^{2}(G_{S}-4G_{V})
\,,
\label{new_rel}
\end{eqnarray}
where ${\bar M}_{A}^{2} = M_{A}^{2}/(e_{\Lambda}^{2} \Lambda^{2})$.

\section{Summary and discussion}

In the ERG with a momentum cutoff, the flow equation generates gauge 
non-invariant quantum corrections such as a photon mass term. 
BRST transformation of these symmetry breaking corrections 
are systematically cancelled, if the QME is fulfilled. Using 1PI formulation, 
we have shown that the perturbative solutions to flow equation also 
solve the QME/mST for QED with four-fermi interactions in a general covariant gauge.

As for $\Lambda$ evolution, we obtain the standard anomalous dimensions
and the standard beta function of the gauge coupling together with
$Z_{1} = Z_{2}$ relation when removing four-fermi interactions.  This
reflects regularization-scheme independence in the one-loop computation
for these objects.  When included, the four-fermi terms yields a new
contribution to the beta function. Its coefficient depends on the choice
of cutoff function, and expressed in terms of the photon mass term. Even
for $Z_{1} \neq Z_{2}$, BRST symmetry is unbroken because the QME/mST
remains intact for the rescaled 1PI action (\ref{QED_Gamma012_Z}). It is
a consequence of invariance of QME/mST under a canonical transformation
used in introducing wavefunction renormalization factors.

The use of such a canonical transformation will induce an undesirable
$\Lambda$ evolution in $eZ_{e}Z_{3}^{1/2}\Psi^{*}\Psi C$ and
$eZ_{e}Z_{3}^{1/2}{\bar\Psi}^{*} {\bar\Psi} C$ vertices for $Z_{1} \neq
Z_{2}$, i.e.  $Z_{e}Z_{3}^{1/2} \neq 1$.  Hence, RG flows should be
computed suppressing the antifields at the final stage. Then,
(\ref{flow_Sigma}) is satisfied, and RG flows stay in BRST invariant
submanifold of the theory space.

Let us consider the cutoff removing limit $K \to 0~(\Lambda \to 0)$. 
In the one-loop formula (\ref{Str_log_formula}), the 
IR-regulated propagators ${\bar \Delta}^{AB}$ are replaced with unregularized 
ones $\Delta^{AB}$.  The quantum actions $\Gamma_{2,q}$ and $\Gamma_{3,q}$ 
whose UV divergences are removed using the dimensional regularization 
satisfy the Zinn-Justin equations: $(\Gamma_{0}, \Gamma_{2,q})=0$, 
$(\Gamma_{0}, \Gamma_{3,q}) + (\Gamma_{1}, \Gamma_{2,q})=0$. 
The first equation leads to vanishing quantum corrections in the 
longitudinal part of the photon two-point functions, ${\cal L} =0$.
The second equation gives the standard relation between the mass operator of 
the fermion two-point functions and the gauge interaction vertices.
Since the four-fermi interactions yield no contribution 
to $\beta_{e}$ as seen from (\ref{new_rel}) with $M_{A} =0$, 
they do not affect the WT relation $Z_{1} = Z_{2}$.  
Therefore, we observe that the classical BRST symmetry  
persists at quantum level, irrespective of the 
presence of higher dimensional operators such as the four-fermi interactions.
These results should be compared with those for $\Lambda \neq 0$. 

Our perturbative results imply that a non-perturbative study
of the chiral invariant QED will certainly observe a similar modification of the
Ward identity if we include the four fermi interactions.

\section*{Acknowledgment}

This work was supported by JSPS KAKENHI Grant Number 19K03822.



\section*{Appendix A}

To calculate r.h.s. of \bref{QME_Gamma} and \bref{flow_Gamma}, we need
to find field dependent parts of $\Gamma_{I}^{(2)}$ and
$\Gamma_{I*}^{(2)}$, the vertices denoted as $\tau$.

From $\Gamma_{1}$ we find the following vertices,
\begin{eqnarray}
&& 
\tau_{\hat\al\beta}^{(-\Slash{A})}(x,y) = 
\frac{\pa^{l} \pa^{r} \Gamma_{1}}{\pa \bar\Psi_{\hat\al}
(x) \pa \Psi_{\beta}(y)}
= -e  (\Slash{A})_{\hat\al\beta}(x) \delta(x-y) \,,
\nn\\
&& 
\tau_{\al\hat\beta}^{(\Slash{A}^{T})}(x,y) = 
\frac{\pa^{l} \pa^{r} \Gamma_{1}}{\pa \Psi_{\al}(x) 
\pa \bar\Psi_{\hat\beta}(y)}
= + e  (\Slash{A}^{T})_{\al\hat\beta}(x) \delta(x-y) \,,
\label{vertices}
\\
&& 
\tau_{\hat\al\mu}^{(-\gamma\Psi)}(x,y) = 
\frac{\pa^{l} \pa^{r} \Gamma_{1}}{\pa \bar\Psi_{\hat\al}(x) \pa A_{\mu}(y)} 
= - e (\gamma_{\mu}\Psi)_{\hat\al}(x) \delta(x-y)
=  -\tau_{\mu\hat\al}^{(\gamma\Psi)}(x,y) \,,
\nn\\
&& 
\tau_{\mu\beta}^{(-\bar\Psi\gamma)}(x,y) = 
  \frac{\pa^{l} \pa^{r} \Gamma_{1}}{\pa 
A_{\mu}(x) \pa\Psi_{\beta}(y)} = - e (\bar\Psi\gamma_{\mu})_{\beta}(x) 
\delta(x-y) = -\tau_{\beta\mu}^{(\bar\Psi\gamma)}(x,y)  \,.
\nn
\end{eqnarray}
Here the superscripts of $\tau$ indicate structures of vertices. 
Similarly from $\Gamma_{2,{\rm cl}}^{(2)}$, we have
\begin{eqnarray}
\tau_{\hat\al\beta}^{(\bar\Psi\Psi)}(x,y) 
&=& \left[
\frac{\pa^{l} \pa^{r} \Gamma_{2,{\rm cl}}[\Phi]}{\pa \bar\Psi_{\hat\al}
(x) \pa \Psi_{\beta}(y)}\right]_{A=0} 
\nn\\
&=& G_{S}\delta(x-y)\biggl\{\Bigl[
 \delta_{{\hat\al}\beta}\bigl({\bar\Psi}(x)\Psi(x) \bigr)
- (\gamma_{5})_{{\hat\al}\beta} \bigl({\bar\Psi}(x)\gamma_{5}
\Psi(x) \bigr)\Bigr] 
\nn\\
&& 
- \Bigl[{\bar\Psi}_{\beta}(x)\Psi_{\hat\al}(x) 
- ({\bar\Psi}\gamma_{5})_{\beta}(x)(\gamma_{5}\Psi)_{\hat\al}(x)
\Bigr]
\biggr\}
\label{4f_2}\\
&& + G_{V}\delta(x-y) \biggl\{
\Bigl[
(\gamma_{\mu})_{{\hat\al}\beta}\bigl({\bar\Psi}(x)
\gamma_{\mu}\Psi(x)\bigr) 
+(\gamma_{5}\gamma_{\mu})_{{\hat\al}\beta}
\bigl({\bar\Psi}(x)\gamma_{5}\gamma_{\mu}\Psi(x)\bigr)\Bigr]
\nn\\
&& 
- \Bigl[\bigl({\bar\Psi}(x)\gamma_{\mu})_{\beta}
(\gamma_{\mu}\Psi(x)\bigr)_{\hat\al} 
+ \bigl({\bar\Psi}(x)\gamma_{5}\gamma_{\mu})_{\beta}
(\gamma_{5}\gamma_{\mu}\Psi(x)\bigr)_{\hat\al}\Bigr]\biggr\}
\nn\\
&=& - \bigl(\tau^{(\bar\Psi\Psi)}\bigr)^{T}_{\beta\hat\al}(y,x)\,.
\nn
\end{eqnarray}

As for $\Gamma_{I*}^{(2)}$, we need vertices out of $\Gamma_1$,
\begin{eqnarray}
&& 
\tau^{C}_{*\al\beta} = \frac{\pa}{\pa \Psi_{\al}^{*}(x)}
\frac{\pa^{r}}{\pa \Psi_{\beta}(y)} \Gamma_{1} = + ie \delta_{\al\beta} 
C(x) \delta(x-y)  \,,
\nn\\
&& 
\tau^{C}_{*\hat\al\hat\beta} = \frac{\pa}{\pa \bar\Psi_{\hat\al}^{*}(x)}
\frac{\pa^{r}}{\pa \bar\Psi_{\hat\beta}(y)} \Gamma_{1} = - ie 
\delta_{\hat\al\hat\beta} 
C(x) \delta(x-y)  \,.
\label{Gamma*}
\end{eqnarray}

\section*{Appendix B}

The following example shows our notation for computing the QMF.
\begin{eqnarray} 
{\rm Tr} \Bigl[K  C \Slash{\pa}{\mathbf {\bar\Delta}}  \Slash{A} \Bigr] 
&=& \int_{x,y} K(x-y) C(y) {\rm tr}\Bigl[
\Slash{\pa}_{y}{\mathbf {\bar\Delta}}(y-x) \Slash{A}(x)\Bigr] \nn\\
&=&\int_{x,y}{\rm tr}\Bigl[\Slash{A}^{T}(x)
\Slash{\pa}^{T}_{x}{\mathbf {\bar\Delta}}(y-x)\Bigr]C(y)K(x-y) 
\nn\\
&=& - \int_{x,y}{\rm tr}\Bigl[\Slash{A}(x)
\Slash{\pa}_{x}{\mathbf {\bar\Delta}}(x-y)\Bigr]C(y)K(y-x)
\nn\\
&=& - \Bigl[\Slash{A}\Slash{\pa}{\mathbf {\bar\Delta}}C K\Bigr]\,,
\label{B1}
\end{eqnarray}
where the trace is taken for $\gamma$ matrices. We have also used the
charge conjugation relation $C \gamma_{\mu}^{T} C^{-1} = -
\gamma_{\mu}$, and symmetry properties ${\mathbf {\bar\Delta}}(x-y) =
{\mathbf {\bar\Delta}}(y-x)$ and $K(x-y) = K(y-x)$.

In this notation, we obtain eq. \bref{Sigma_AC} by making integration 
by parts as 
\begin{eqnarray}
&&\Sigma_{2}|_{(\Gamma_{0}, \Gamma_{2,q}) } = - \frac{e^{2}}{2}\Bigl[
\Slash{\pa}{\mathbf {\bar\Delta}}\Slash{A}\Slash{\pa}{\mathbf {\bar\Delta}}
\Slash{\pa}C + \Slash{\pa}{\mathbf {\bar\Delta}}\Slash{\pa}C
\Slash{\pa}{\mathbf {\bar\Delta}}\Slash{A}
\Bigr] = - e^{2}\Bigl[\Slash{\pa}{\mathbf {\bar\Delta}}\Slash{\pa}C
\Slash{\pa}{\mathbf {\bar\Delta}}\Slash{A}\Bigr]
\nn\\
&&~= e^{2} \int_{x,y}
{\rm tr}\Bigl[\Slash{\pa}_{x}\Slash{\pa}_{y}{\mathbf {\bar\Delta}}(x-y)C(y)
\Slash{\pa}_{y}{\mathbf {\bar\Delta}}(y-x)\Slash{A}(x)
+ \Slash{\pa}_{x}{\mathbf {\bar\Delta}}(x-y)C(y)
\Slash{\pa}_{y}^{2}{\mathbf {\bar\Delta}}(y-x)\Slash{A}(x)
\Bigr]
\nn\\
&&~= e^{2} \Bigl[(1-K)C\Slash{\pa}{\mathbf {\bar\Delta}}\Slash{A}
- \Slash{\pa}{\mathbf {\bar\Delta}}C (1-K)\Slash{A}\Bigr]
= 2e^{2}\Bigl[(1-K)C\Slash{\pa}{\mathbf {\bar\Delta}}\Slash{A}\Bigr]
\label{B2}
\end{eqnarray}

\end{document}